\documentclass[aps,prd,superscriptaddress,10pt,twocolumn]{revtex4}
\usepackage{graphicx}
\usepackage{epstopdf}
\usepackage{amsmath}
\usepackage{amsfonts}
\usepackage{amssymb}
\usepackage{latexsym}
\usepackage{hyperref}
\usepackage[english]{babel}
\usepackage[utf8]{inputenc}
\usepackage[colorinlistoftodos]{todonotes}
\usepackage{xcolor}
\usepackage{slashed}
\usepackage{bm}  
\usepackage{subcaption}
\usepackage[normalem]{ulem} 
\usepackage{amsmath,mathtools}
\usepackage{caption}
\usepackage{subcaption}
\usepackage{float}
\useunder{\uline}{\ul}{}
\usepackage{hyperref}

\usepackage{layout}
\begin{document}

\title{Limits on non-minimal Lorentz violating parameters through FCNC and LFV processes}

\author{Y. M. P. Gomes}\email{ymuller@cbpf.br}
\affiliation{Centro Brasileiro de Pesquisas F\'{i}sicas (CBPF), Rua Dr Xavier Sigaud 150, Urca, Rio de Janeiro, Brazil, CEP 22290-180}

\author{J. A. Helayel-Neto}\email{helayel@cbpf.br}
\affiliation{Centro Brasileiro de Pesquisas F\'{i}sicas (CBPF), Rua Dr Xavier Sigaud 150, Urca, Rio de Janeiro, Brazil, CEP 22290-180}


\begin{abstract}
In this work we analyze a non-minimal Lorentz-violating extension of the electroweak theory in the fermionic sector. Firstly we analyze the relation between the CKM rotation in the quark sector and possible contributions of this new coupling to flavor changing neutral currents (FCNC) processes. In sequel we look for non-diagonal terms through possible lepton flavor violation (LFV) decays. Strong bounds are presented to the Lorentz violating parameters of both the quark and the lepton sectors.     
\end{abstract}

\maketitle

\pagestyle{myheadings}
%

\section{Introduction} \label{sec_intro}

\paragraph*{}Despite its great success, the Standard Model of Particle Physics should not be the final description of nature and it has been shown that in some of its extensions, string theory, for example, is possible that Lorentz symmetry is violated \cite{intro1, intro2}. Observation of any, albeit small, sign of Lorentz symmetry violation (LSV) would represent a major paradigm shift and would require re-examination of the very basis of modern physics, i.e., relativity theory and quantum field theory \cite{intro1,intro2, intro3, intro4}.

\paragraph*{}A possible realization of LSV is achieved with Lagrangian model where a spin field acquires a non-zero vacuum expectation value - see for example, Ref. \cite{intro1}. Given this approach, one can introduce non-dynamic tensors \cite{intro5} and explore various differences between couplings for the SM  gauge and matter sectors \cite{intro1, intro2, intro3, intro4, intro5, intro6}. For a review of theory and experimental tests of  CPT and Lorentz invariance, see references \cite{intro4,intro5,intro6, intro7,intro8,intro9,intro10,intro11}.
\paragraph*{}In the present study we investigate the case proposed in \cite{elweakLSV1} of a non-minimally coupling with a constant 4-vector background with a specific SM sector, the fermion - electroweak Bosons interacting sector. More specifically, we follow two main paths which are; possible contributions for flavor Changing Neutral Currents (FCNC) decays and contributions for Lepton flavor Violation (LFV) decays \cite{SM1,MEG}. Studies involving mesons and LSV can be seen in \cite{kostmesons,bsmesonLHC}. In the first path we use the strong bounds in the FCNC mesons decays given by $K^0 \rightarrow \mu^+ \mu^-$, $B^0 \rightarrow \mu^+ \mu^-$ and $B^0_s \rightarrow \mu^+ \mu^-$ to find constrains involving the LSV parameters. Going further, we analyze the lepton sector through the LFV decays $\mu \rightarrow e+ \gamma$, $\tau \rightarrow \mu + \gamma$ and $\tau \rightarrow e+ \gamma$. In the final comments we discuss our results.  

\section{Non-minimal coupling in the Electroweak Sector}

\paragraph*{} Based on the non-minimal coupling used in \cite{elweakLSV1} one can extend the idea of a non-minimal coupling in the $SU(2) \times U(1) $ sector of the Standard Model. Starting from the implementation of the following covariant derivative proposed we have 
\begin{eqnarray}\nonumber
(D_\mu')_{AB} &=& \Big{(} \partial_\mu + i g Y B_\mu + i g' W_\mu^I \sigma^I/2 \Big{)}\delta_{AB} + \\&&- i \xi^\nu_{AB} F_{\mu \nu} - i \rho^\nu_{AB} \sigma^I F^I_{\mu \nu}~,
\end{eqnarray}
 where $\xi^\nu_{AB}$ and $\rho^\nu_{AB}$ are LSV parameters, $Y = 2(Q - T^3)$, $g$ and $g'$ are the $U(1)_Y$ and $SU(2)_L$ coupling constants, respectively, $B_\mu$ and $W_\mu^I$ are the $U(1)_Y$ and $SU(2)_L$ Gauge bosons, the indexes $ A, B $ refers to Standard Model fermionic families, $F_{\mu \nu} = \partial_{[\mu} B_{\nu]}$, {\small \begin{equation}\nonumber 
F_{\mu \nu}^I \sigma^I =  \begin{pmatrix}
    \partial_{[\mu} W_{\nu]}^3 + g W^+_{[\mu} W^-_{\nu]}      & \partial_{[\mu} W_{\nu]}^+ + g W^3_{[\mu} W_{\nu]}^+  \\
    \partial_{[\mu} W^-_{\nu]} + g W^3_{[\mu} W_{\nu]}^-       & - \partial_{[\mu} W^3_{\nu]} -g W^+_{[\mu} W_{\nu]}^-   \\ 
\end{pmatrix} ~, 
\end{equation} } $\sigma^I$ refers to Pauli's matrices. Our analysis will begin in the quarks sector, where we expect to analyze the relationship between the Lorentz violation and the CP violation that is represented by the CKM Matrix. In the sequence we will analyze the lepton sector, where the absence of Right neutrinos and the universality of weak interactions in this sector demand a thorough analysis.

\subsection{The Quark sector and FCNC:}
Taking into account the flavor structure, the new covariant derivative will act in the quarks sector as follows:
\begin{eqnarray}\nonumber
\mathcal{L}&=&  (\bar{Q}_L)_A(i \gamma^\mu D_\mu' )_{AB} (Q_L)_B +\\\nonumber  &&+ (\bar{u}_R)_A(i \gamma^\mu D_\mu' )_{AB} (u_R)_B +\\  &&+ (\bar{d}_R)_A(i \gamma^\mu D_\mu' )_{AB} (d_R)_B   ~,
\end{eqnarray}
where $ A,B $ refers to the respective quark family (i.e., flavor). Here we use $(Q_L)_A = (u_{L,A} ~ d_{L,A})^T$ doublet under $SU(2)_L$ and $(u_R)_A$, $(d_R)_A$ singlets under $SU(2)_L$. The Lagrangian above brings us the following new interaction terms in the Left sector:
\begin{eqnarray}\nonumber \label{quarklv1}
\mathcal{L}_{LSV}^{Left}&=& \xi^\mu_{AB} (\bar{Q}_L)_A \gamma^\nu (Q_L)_B F_{\mu \nu} +\\
&&+ \rho^\mu_{AB} (\bar{Q}_L)_A \gamma^\nu \sigma^I (Q_L)_B F^I_{\mu \nu}~, 
\end{eqnarray}

Since this coupling acts in the interaction sector of the fermionic sector with the electroweak bosonic fields, there will be no changes in the fermion masses neither in the gauge boson mass. As we know, the relationship between the bosons $ B, W^3 $ and $ A, Z $ is given from the mass matrix diagonalization after the spontaneous symmetry breaking of the Higgs potential, and from this diagonalization appears the very known relationship:
\begin{equation}
    A = c_W B + s_W W^3 ~~, ~~ Z = - s_W B + c_W W^3~.
\end{equation}

where $c_W = \cos \theta_W$, $s_W =\sin \theta_W$ and $\theta_W$ is the Weinberg angle. Rewriting the Lagrangian \eqref{quarklv1} in terms of the photon and Z-boson fields, we reach:
\begin{eqnarray}\nonumber
\mathcal{L}^{Left}_{LSV} &=& \bar{u}_{L,A}( \Delta_{AB} +\tilde{\Delta}_{AB}) u_{L.B} \\\nonumber &&+ \bar{d}_{L,A}(  \Delta_{AB} - \tilde{\Delta}_{AB})d_{L,B}  \\ &&+ \bar{u}_{L,A} \Delta_{AB}^+  d_{L,B}  + h.c.~~,
\end{eqnarray}

where
\begin{equation}
     \Delta_{AB} = \xi^{[\mu}_{AB} \gamma^{\nu]} (c_W \partial_\mu A_\nu - s_W \partial_\mu Z_\nu) ~~,
\end{equation}

\begin{equation}
     \tilde{\Delta}_{AB} = \rho^{[\mu}_{AB} \gamma^{\nu]} (s_W \partial_\mu A_\nu + c_W \partial_\mu Z_\nu + i g' W^+_\mu W^-_\nu)~, 
\end{equation}
and
\begin{equation}
     \Delta_{AB}^+ = \rho^{[\mu}_{AB} \gamma^{\nu]} ( \partial_\mu + i g' s_W A_\mu  + i g' c_W  Z_\mu )W^+_\nu~.
\end{equation}
\paragraph*{}Similarly, the couplings in the Right sector gives us the following interaction Lagrangian:
\begin{equation}
    \mathcal{L}_{LSV}^{Right} = \xi^\mu_{AB}\left( \bar{u}_{R,A} \gamma^\nu u_{R,B}  +  \bar{d}_{R,A} \gamma^\nu d_{R,B} \right)F_{\mu \nu}~.   
\end{equation}
\paragraph*{}The equation above is justified by the fact that that the Right sector is singlet under $ SU (2)_L $ group transformations. Rewriting the above equation on the same $\{A,Z\}$ basis we have:
\begin{equation}
    \mathcal{L}_{LSV}^{Right} =  \bar{u}_{R,A} \Delta_{AB} u_{R,B} +  \bar{d}_{R,A}\Delta_{AB} d_{R,B} ~~. 
\end{equation}
  
\paragraph*{}
Finally, introducing the CKM matrix we have that with the standard parametrization choice, $d'_A = V_{AB} d_B $ where $d'_A$ represent the physical quarks. The CKM matrix expansion in terms of standard  parameters $\lambda$, $A$, $\rho$ e $\eta$ ($\lambda \approx | V_{us} | \approx 0.23$, where $ s_{12} = \lambda $, $ s_{23} = A \lambda^2 $, $s_{13} e^{-i\delta} = A\lambda^3(\rho - i \eta)$), with $
A \approx 0.83$, $\rho \approx 0.12$ and $\eta \approx 0.35$, gives the following CKM matrix approximation \cite{SM4}:
\begin{widetext}

\begin{equation}
V_{CKM} \approx \left(
\begin{array}{ccc}
 1-\frac{\lambda ^2}{2}-\frac{\lambda ^4}{8} & \lambda  & A \lambda ^3 (\rho -i \eta ) \\
 -\lambda  & 1-\frac{\lambda ^2}{2}-\frac{1}{8} \left(4 A^2+1\right) \lambda ^4 & A \lambda ^2 \\
 A \lambda ^3 (1 -\rho -i \eta) & -A \lambda ^2+\frac{1}{2} A \lambda ^4 (1-2 (\rho -i \eta )) & 1-\frac{A^2 \lambda ^4}{2} \\
\end{array}
\right)~~.
\end{equation}
\end{widetext}
\paragraph*{}So we can rewrite the new Lagrangian and interaction with the Left and Right sectors. $\mathcal{L}_{LSV}^{total} = \mathcal{L}_{LSV}^{Left} + \mathcal{L}_{LSV}^{Right}$  and it is given by:
\begin{eqnarray}\nonumber
  \mathcal{L}_{LSV}^{total} &=&  \bar{U}_{A}( \Delta_{AB} +\tilde{\Delta}_{AB}P_L) U_{B} +\\\nonumber &&+ \bar{D}_{A}V_{AC}(  \Delta_{CD} - \tilde{\Delta}_{CD}P_L)V^\dagger_{DE}D_{E}  +\\ && + \bar{U}_{i} \Delta_{AB}^+ V_{BC}^\dagger P_L D_{C}  + h.c. ~~.
\end{eqnarray} 
Here $U_A$ , $D_A$ are Dirac spinors $U_A = (u_{L,A} ~ u_{R,A})^T$, $D_A = (d_{L,A}' ~ d_{R,A}')^T$ and $P_L = \frac{(1- \gamma_5)}{2} $. The Interaction Lagrangian can be written explicitly as follows:
\begin{eqnarray}\nonumber\label{eqfinalFCNC}
\mathcal{L}_{LSV}^{total} &=&  \overline{U}_A  \left( c_1^\mu \gamma^\nu + c_2^\mu \gamma^\nu \gamma_5  \right)_{AB}  U_B  A_{\mu\nu} 
+\\\nonumber  &&+\overline{D}_{A} \left( c_3^\mu \gamma^\nu + c_4^\mu \gamma^\nu \gamma_5 \right)_{AB} D_{B}  A_{\mu\nu} +
 \\ \nonumber
&&+  \overline{U}_A  \left( c_5^\mu  \gamma^\nu + c_6^\mu \gamma^\nu \gamma_5 \right)_{AB} U_B  Z_{\mu\nu} +\\\nonumber &&
+  \bar{D}_{A} \left( c_7^\mu \gamma^\nu + c_8^\mu \gamma^\nu \gamma_5 \right)_{AB} D_B Z_{\mu\nu} + \\\nonumber
&& + \frac{1}{2} (c_9^\mu)_{AB}  \overline{U}_A \gamma^\nu \left( 1-\gamma_5 \right) D_B  \nabla_{[\mu} W^-_{\nu]}+ \\\nonumber
&&+ \frac{i}{2} g  W^+_{[\mu} W^-_{\nu]} \big{[}(\rho^\mu)_{AB} \overline{U}_A \gamma^\nu \left(1-\gamma_5\right) U_B -\\\nonumber
&&~~-(\hat{\rho}^\mu)_{AB} \bar{D}_{A} \gamma^\nu \left( 1-\gamma_{5} \right) D_B  \big{]}+ \text{h.c.} \; ~,\\
\end{eqnarray}
where $A_{\mu \nu} = \partial_{[\mu} A_{\nu]}$, $Z_{\mu \nu} = \partial_{[\mu} Z_{\nu]}$, $\nabla_\mu W^-_\nu =  \partial_{[\mu} W_{\nu]}^- +i e A_{[\mu}W_{\nu]}^- + i e \cot \theta_W Z_{[\mu}W_{\nu]}^- $, $\hat{n}_{AB} = (V n V^\dagger)_{AB}$ for any matrix  $n_{AB}$, $c_1^\mu = \frac{1}{2}c_W \xi^\mu + \frac{1}{4}s_W \rho^\mu$,$ c_2^\mu = - \frac{1}{4}s_W \rho^\mu$, $c_3^\mu =  \frac{1}{2}c_W \hat{\xi}^\mu - \frac{1}{4}s_W \hat{\rho}^\mu$, $c_4^\mu = \frac{1}{4} s_W \hat{\rho}^\mu$, $c_5^\mu = - \frac{1}{2}s_W \xi^\mu - \frac{1}{4} c_W \rho^\mu$, $c_6^\mu = \frac{1}{4} c_W \rho^\mu$, $c_7 = - \frac{1}{2}s_w \hat{\xi}^\mu + \frac{1}{4}c_W \hat{\rho}^\mu$, $c_8^\mu = \frac{1}{4} c_W \hat{\rho}^\mu$, $c_9 = \rho^\mu V^\dagger$, $s_W = \sin \theta_W$, $c_W = \cos \theta_W$, and we omit flavor indexes for simplicity. We shall pay attention to $c_8$ parameter, since it will be the main parameter in which contributes to the process we are interested. The set of vertex are shown in table \ref{vertexLVFCNC}.  

\renewcommand{\arraystretch}{1.8}
\begin{table}[htb!]
\centering
\begin{tabular}{|c|c|}
\hline
\quad  \quad {\bf Interaction}    \quad \quad & \quad    {\bf Vertex}    \quad \\
\hline
\hline

 \quad $\gamma \, U_A \, U_B$  \quad  &  \quad $q_\nu ( c_1^{[\mu} \gamma^{\nu]} + c_2^{[\mu} \gamma^{\nu]}\gamma_5 )_{AB}  $  \quad  \\
\hline

 \quad $\gamma \, D_{A} \, D_{B}$ \quad & \quad $q_\nu ( c_3^{[\mu} \gamma^{\nu]} + c_4^{[\mu} \gamma^{\nu]}\gamma_5 )_{AB}  $  \quad \\

\hline

 \quad $Z^0 \, U_A \, U_B$  \quad  &   \quad $q_\nu ( c_5^{[\mu} \gamma^{\nu]} + c_6^{[\mu} \gamma^{\nu]}\gamma_5 )_{AB} $  \quad  \\

\hline

\quad $Z^{0} \, D_{A} \, D_{B}$  \quad  &  \quad $q_\nu ( c_7^{[\mu} \gamma^{\nu]} + c_8^{[\mu} \gamma^{\nu]}\gamma_5 )_{AB}  $  \quad \\

\hline

 \quad $W^{-} \, U_A \, D_{B}$  \quad  &  \quad $\frac{1}{2} q_\nu (c_9)_{AB}^{[\nu} \gamma^{\mu]} \frac{\left(1-\gamma_5\right)}{2} $  \quad \\

\hline

 \quad $W^{-} \, \gamma \, U_A \, D_{B}$\quad &  \quad $\frac{i e}{2} (c_9)_{AB}^{[\nu} \gamma^{\mu]} \frac{\left(1-\gamma_5\right)}{2}$ \quad  \\

\hline

 \quad $W^{-} \, Z^0 \, U_A \, D_{B}$  \quad &  \quad $\frac{i e}{2}\cot \theta_W (c_9)_{AB}^{[\nu} \gamma^{\mu]} \frac{\left(1-\gamma_5\right)}{2}$ \quad \\

\hline

 \quad $W^{+} \, W^{-} \, U_A \, U_B$ \quad \quad & \quad \quad $\frac{i g}{2} (\rho)_{AB}^{[\nu} \gamma^{\mu]} \frac{\left(1-\gamma_5\right)}{2}$ \quad \quad \\

\hline

 \quad $W^{+} \, W^{-} \, D_{A} \, D_{B}$  \quad &  \quad $-\frac{i g}{2} (\tilde{\rho})_{AB}^{[\nu} \gamma^{\mu]} \frac{\left(1-\gamma_5\right)}{2}$ \quad  \\

\hline
\end{tabular}
\caption{ Vertex factors obtained from Eq.\eqref{eqfinalFCNC}. Here $q^\mu$ represents the  $A$, $W$ or $Z$ 4-momentum. } 
\label{vertexLVFCNC}
\end{table}

\paragraph*{}From the equation \eqref{eqfinal}, we can see that the CKM Matrix will not only appear in the charged currents as in the Standard Model, but now we will also get Lorentz violation parameters which can be complex due to the CP violation parametrized by the $ \delta $ complex phase. 

We assume Lorentz violation vectors such that families do not mix, but show dependence on the family. In other words, let's say $ \xi^\mu_{AB} $ is given by:
\begin{equation}\label{sup1}
\xi^\mu_{AB} = \left(
\begin{array}{ccc}
 \text{$\xi^\mu_{11}$} & \text{$0 $} & \text{$0 $} \\
 \text{$ 0$} & \text{$\xi^\mu_{22} $} & \text{$0 $} \\
 \text{$0$} & \text{$0$} & \text{$\xi^\mu_{33} $} \\
\end{array}
\right).
\end{equation}
where $\xi_{11}^\mu \neq \xi_{22}^\mu \neq \xi_{33}^\mu$, in principle. We assume that $ \rho $ shares the same characteristics.  This approximation is based on the fact that non-diagonal elements will allow $\Delta C$ decays ($c \rightarrow u + \gamma/Z$ for instance), highly suppressed in the SM. From this structure, after the rotation induced by the CKM matrix, we obtain:

\begin{equation}
\hat{\xi}_{11} = (V \xi V^\dagger)_{11} \approx \xi_{11} +(\xi_{22}- \xi_{11})\lambda ^2+ O(\lambda^5)  ~,
\end{equation}
\begin{equation}
\hat{\xi}_{12} \approx (-\lambda +\frac{\lambda ^3}{2})(\xi_{11}-\xi_{22})+ O(\lambda^5)  ~,
\end{equation}
\begin{equation}
\hat{\xi}_{13} \approx  (\xi_{11}-\xi_{22})(A \lambda ^3) - (\xi_{11}- \xi_{33})  A \lambda ^3  (\rho - i \eta ) + O(\lambda^5) ~,
\end{equation}
\begin{equation}
\hat{\xi}_{22} \approx \xi_{22} +(\xi_{11}- \xi_{22})\lambda ^2+ (\xi_{33}-\xi_{22})A^2 \lambda ^4 +O(\lambda^5)  ~,
\end{equation}
\begin{eqnarray}\nonumber
\hat{\xi}_{23} &\approx& (\xi_{33}- \xi_{22})A \lambda ^2+\xi_{11}A \lambda ^4 (-i \eta +\rho -1) +\\
&&+\xi_{22}A \lambda ^4 (-i  \eta - \rho +1)  + O(\lambda^5)  ~,
\end{eqnarray}
\begin{equation}
\hat{\xi}_{33} \approx \xi_{33} +(\xi_{22}- \xi_{33})   A^2 \lambda^4  + O(\lambda^5)  ~,
\end{equation}

\paragraph*{}and $\hat{\xi}_{ji} = \hat{\xi}_{ij}^*$ for $i \neq j$. Note that the complex characteristic resides in the non-diagonal components of Lorentz violation parameters, but depends on the difference in magnitudes expressed in our assumption \eqref{sup1}. In the low energy limit, that is, in the limit where the massive gauge bosons decay fast enough, the $ Z $ decay generates the following new effective interaction between neutral currents in the $D$-quark sector:
\begin{eqnarray}\label{FCNC1}\nonumber
\mathcal{H} &=&   \frac{c_W^2 G_F}{\sqrt2}J_\mu^0 \times  \bar{D}_A \gamma^{[\nu} q^{\mu]}(c_{7,\nu} + c_{8,\nu} \gamma_5)_{AB}D_B  ~~,\\
\end{eqnarray}

where $J_\mu^0 = \sum_f\bar{f}  \gamma_\mu(v_f - a_f \gamma_5) f$ in which couples to $Z$ in the SM, $\frac{G_F}{\sqrt2} = \frac{g^2}{8 M_W^2}$, $v_f = T_3^f- 2 Q_f s_W^2$, $a_f = T_3^f$ and $q = p_{\bar{f}}- p_f = p_{D_A}-p_{D_B}$. Interestingly, while $ c_5 $ and $ c_6 $ remain diagonal in flavor space, $ c_7 $ and $ c_8 $ contain contributions from the CKM matrix, so they will not be diagonal, assuming Eq. \eqref{sup1}. The $ c_3 $ and $ c_4 $ parameters also share this property. Importantly, the terms $ c_7 $ and $ c_8 $ generate the so-called flavor Changing Neutral Current (FCNC) in a tree-level process. In SM this processes are forbidden in a tree level process, so it is an important result to be discussed. In addition, it is also important to point out that $ c_3 $ and $ c_4 $ can generate flavor exchange through photon coupling, a phenomenon that does not occur in the SM.
\paragraph*{}Analyzing in detail the interaction that brings us the possibility of FCNC processes given by the equation \eqref{FCNC1}, after some algebraic manipulations we can rewrite it as follows:
\begin{eqnarray}\label{FCNC2}\nonumber
\mathcal{H}^{FCNC} &=& \frac{c_W^2 G_F}{\sqrt2}\Big{[} \bar{D}_A \gamma^\mu D_B  M_{\mu ~ AB}^{~\nu}  J_\nu^0 +\\\nonumber &&+\bar{D}_A \gamma^\mu \gamma_5 D_B   N_{\mu ~ AB}^{~\nu} J_\nu^0   \Big{]}~~,
\end{eqnarray}

where $M_{\mu ~AB}^{~\nu} = (\delta_\mu^{\nu} q.c_7 - q_\mu c_{7}^\nu )_{AB}$ and $N_{\mu ~ AB}^{~\nu} = ( \delta_\mu^{\nu} q.c_8 - q_\mu c_{8}^\nu )_{AB}$. Now we can calculate the Decay Rate of some possible processes in the non-diagonal sector in the flavor space. 

\begin{figure}[H]
\centering
\includegraphics[height=2cm, angle = 0]{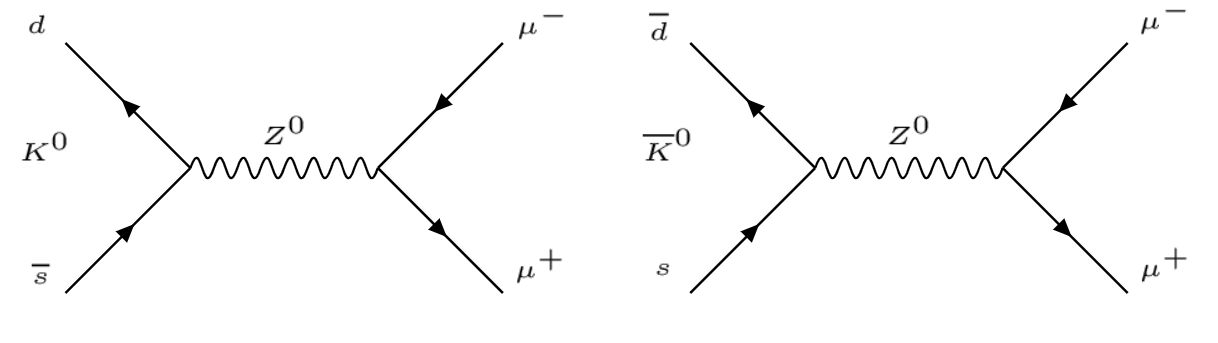}
\caption{ Hypothetical decay of the neutral Kaons $ K ^ 0 $ and $ \bar{K}^0 $ in a FCNC process.}
\label{Kdecay}
\end{figure}

\paragraph*{}Neutral Kaon Decay: If there is a contribution to the neutral current with FCNC then the neutral K mesons ($ \bar {s} d $ and $\bar{d} s$ ) could easily decay into a pair of muons, as is shown in Fig. \ref{Kdecay}.

The standard method to work in meson decays is given by the following parametrization $\langle 0| \bar{s} \gamma_\mu \gamma_5 d | K^0(q)\rangle = i F_{K^0} q_\mu $ \cite{elweakLSV1}. The vector current will be responsible for the vector mesons (e.g. , $K^*$), and a similar analysis can be done. Thus, the scattering matrix of the Kaon decay is given by: 
\begin{eqnarray}\nonumber
\mathcal{M} &=& \frac{c_W^2 G_F}{\sqrt2}(N_{12})^{\mu \nu}\times \langle 0| \bar{s} \gamma_\mu \gamma_5 b | K^0(q)\rangle \times \\
&&\bar{\mu}(p)\gamma_\nu (v_f - a_f \gamma_5) \mu(q-p) ~~,
\end{eqnarray}
where $( N)^{\mu \nu}_{~~ 12} = (\eta^{\mu \nu} q.(c_8)_{12} - q^\mu (c_8)_{12}^\nu)$. Therefore, summing over spins of the square of the scattering matrix we reach:
\begin{eqnarray}\nonumber
\langle |\mathcal{M} |^2 \rangle = \frac{c_W^4 G_F^2}{2}F_K^2 q_\mu q_\nu N^{\mu \kappa}_{~~ 12}( N^\dagger)^{\nu \lambda}_{~~ 12}\langle J_\kappa J^\dagger_\lambda \rangle ~,\\
\end{eqnarray}
where $J_\kappa = \bar{\mu}(p)\gamma_\kappa (v_\mu - a_\mu \gamma_5) \mu(q-p)$. After simplifications and using the rest frame of the neutral Kaon where $q = (M_{K^0}, \vec{0})$ we reach :
\begin{eqnarray}\nonumber
\langle |\mathcal{M} |^2 \rangle &\approx & c_W^4 G_F^2 F_{K^0}^2  a_\mu^2 M_{K^0}^6 \Big{(}|~\vec{c}~|^2 - 4 \left(\frac{\vec{c}\cdot \vec{p}}{M_{K^0}} \right)^2 \Big{)} ~,\\&& \hspace{-2.0cm} ~
\end{eqnarray}

where  we ignore terms proportional to $v_\mu = 1- 4 s_W^2  <<1$ and $c = c_{8,12}= \frac{1}{4}c_W \hat{\rho}_{12}$. By the use of the golden rule for two body decay we have:  
\begin{equation}
\Gamma = \frac{1}{2 (4 \pi)^2 M_{K^0}} \int d^3 \vec{p} \frac{\langle |\mathcal{M} |^2 \rangle}{E_p E_{p'}} \delta( M_{K^0} - E_p - E_{p'}) ~~,
\end{equation}

where $p$ and $p'$ are the 4-momentum of the final states. Thus, the integrals are given as follows: 
\begin{eqnarray}\nonumber
\Gamma &=& \frac{c_W^4 G_F^2 F_{K^0}^2  a_\mu^2 M_{K^0}^5}{2 (4 \pi)^2} \int \frac{P d P \delta(P - \kappa)}{2\sqrt{P^2 + y^2 M_{K^0}^2}} \times \\&&\int d\Omega  \left(|~\vec{c}~|^2 - 4 \left(\frac{\vec{c}\cdot \vec{p}}{M_{K^0}} \right)^2\right)~,
\end{eqnarray}
where $P = |\vec{p}|$, $\kappa =  \frac{M_{K^0}}{2} \sqrt{(1-4 y^2)}$ and $y=\frac{m_\mu}{M_{K^0}}$. Defining $\vec{c} = |\vec{c}|( \sin \theta_c \cos\phi _c,\sin \theta_c \sin \phi_c,\cos \theta_c )$ with $\phi_c$, $\theta_c$ generic angles, the angular integral can be calculated and it are given by:
 \begin{equation}
\int d\Omega  \left(|~\vec{c}~|^2 - 4 \left(\frac{\vec{c}\cdot \vec{p}}{M_{K^0}} \right)^2\right) = \frac{4}{3} \pi |~\vec{c}~|^2 (3-\frac{4 P^2}{M_{K^0}^2})~~,
\end{equation}
Therefore:
\begin{eqnarray}\nonumber
\Gamma(K^0 \rightarrow \mu^+ \mu^-) &=& \frac{c_W^4 G_F^2 F_{K^0}^2  a_\mu^2 M_{K^0}^5}{24 \pi } | ~\vec{c}~|^2 \times \\\nonumber &&\hspace{-2.0cm}
\int \frac{P d P}{2\sqrt{P^2 + y^2 M_{K^0}^2}} \delta(P - \kappa)   \left(3-\frac{4 P^2}{M_{K^0}^2}\right) =\\ \nonumber
&&\hspace{-3.0cm}=\frac{c_W^4 G_F^2 F_{K^0}^2  a_\mu^2 M_{K^0}^5}{12 \pi} | ~\vec{c}~|^2 \sqrt{(1-4 y^2)}\left(1 + 2 y^2 \right)~~.\\
\end{eqnarray}
So, applying the value of the constants we found the decay rate as follows:
\begin{equation}
\Gamma(K^0 \rightarrow \mu^- \mu^+) = 4.3 \times 10^{-7} ~|\vec{c}_{12}|^2   MeV^3~~,
\end{equation}
where  $a_\mu \approx 0.5$, $G_F = 1.16 \times 10^{-11} \text{MeV}^{-2}$, $M_{K^0} = 497.61 $ MeV and $F_{K^0} = 164 $ MeV \cite{PDG}. According to \cite{PDG}, in the Standard Model neutral Kaons with long half-lives $ K_L^0 $ have a Decay Rate given by:\begin{equation}
\Gamma(K^0_L) = \frac{1}{\tau_{K^0_L}} \approx 1.3 \times 10^{-14} MeV~~,
\end{equation}
such way that  the Branching Ratio $BR(K_L^0 \rightarrow \mu^+ \mu^-)$, experimentally limited to a value less than $6.8 \times 10^{-9}$ \cite{PDG}, will be given by :

\begin{equation}
BR(K_L^0 \rightarrow \mu^+ \mu^-) = 6.7 \times 10^7 \left( \frac{|\vec{c}|}{MeV^{-1} }  \right) ^2~~.
\end{equation}

Therefore, the contribution to this decay arise from a possible Lorentz violation must be smaller or in the order of contribution from the Standard Model. From this statement we find the following bound for the spacial components of the background vectors:

\begin{equation}
|\vec{\rho}_{22} - \vec{\rho}_{11}| \approx \frac{4}{c_W \lambda} |\vec{c}_{12}| < 5.6 \times 10^{-10} MeV^{-1}~~.
\end{equation}
\paragraph*{}Similarly we can calculate the Branching Ratio for the $B^0$ ($d \bar{b}$) meson, using $M_{B^0}= 5279,6$ MeV, $F_{B^0} = 186 $ MeV \cite{PDG}:

\begin{equation}
BR(B^0 \rightarrow \mu^+ \mu^-) = 1.7 \times 10^8 \left(\frac{|\vec{c}_{13}|}{MeV^{-1}} \right)^2  ~~,
\end{equation}
and using the latest data from \cite{PDG} we reach the following bound:

\begin{eqnarray}\nonumber
|\vec{\rho}_{11}- \vec{\rho}_{22} - (\rho- i \eta)(\vec{\rho}_{11}- \vec{\rho}_{33}) | &\approx& \\\nonumber &&  \hspace{-4.0cm}\approx\frac{4|\vec{c}_{13}|}{A \lambda^3c_W} <  4.2 \times 10^{-8} MeV ^{-1}~.\\
  \end{eqnarray}

\paragraph*{}Finally, for the $B_s^0$ meson ($\bar{s}b$) we find, using $M_{B^0_s}= 5366.9 $ MeV, $F_{B^0_s} = 224$ MeV \cite{PDG}::
\begin{equation}
BR(B_s^0 \rightarrow \mu^+ \mu^-) = 2.7 \times 10^8 \left(\frac{|\vec{c}_{23}|}{MeV^{-1}} \right)^2~~, 
\end{equation}

and therefore we have:
\begin{equation}
|\vec{\rho}_{33} - \vec{\rho}_{22}| \approx \frac{4 |\vec{c}_{23}|}{A \lambda^2 c_W}  < 1.4 \times 10^{-8} MeV^{-1}~~.
\end{equation}

In summary, the bounds obtained are organized in the table \ref{results1}. In the next section we will analyze the lepton sector.

\subsection{The lepton sector and LFV :}
Analyzing now the lepton sector we shall recall that there is no CKM mechanism in this sector, so we shall be presenting an analysis based on an assumption of non-diagonal LSV parameters. As we will see, this non-diagonal terms gives us stronger bounds than the diagonal ones showed in the quark sector. Shallowly speaking,  we have that after modification of the covariant derivative the Lagrangian corresponding to the lepton sector can be written as follows \cite{elweakLSV1,nosso}:
\begin{eqnarray}\label{eq_lag_gl} \nonumber
\mathcal{L}_{\rm \ell} &=& i(\bar{L}_L)_{A} \gamma^\mu (D'_\mu)_{AB} (L_L)_{B}  +\\ && + \, i(\bar{\ell}_R)_{A} \gamma^\mu (D'_\mu)_{AB} (\ell_R)_B\; ,
\end{eqnarray}
here $(L_L)_A = \left( (\nu_L)_A, (\ell_L)_A \right)^T$ and $(\ell_R)_B$ are $SU(2)_L$ doublet and singlet respectively, and mass terms from Yukawa interactions are omitted as LSV terms do not influence them. Lagrangian \eqref{eq_lag_gl} can be split into two components $\mathcal{L}_{ \ell} = \mathcal{L}_{ \ell, SM} + \mathcal{L}_{ LSV}$ and the last component can be written as follows
\begin{eqnarray}\nonumber
\mathcal{L}_{LSV} &=& \frac{1}{2} \, \rho^\mu_{AB} (\bar{L}_{L})_A\gamma^\nu F^I_{\mu\nu} \sigma^I (L_L)_B+\\\nonumber&&\hspace{-1.0cm} +\frac{1}{2} \, \xi^\mu_{AB} \left[ (\bar{L}_{L})_A \gamma^\nu (L_L)_B + (\bar{\ell}_R)_A \gamma^\nu (\ell_R)_B \right]  F_{\mu \nu}~.\\ \hspace{-1.0cm}
\end{eqnarray}

Writing the $ B_\mu $ and $ W_{\mu}^3 $ fields at the base of the $Z$ and the photon fields the above equation will bring us the following Lagrangian interaction for the Left sector:
\begin{eqnarray}\nonumber
\mathcal{L}_{LSV}^{Left} &=& (v_1)^\mu_{AB} (\bar{\ell}_L)_{A} \gamma^\nu (\ell_{L})_{B} \partial_{[\mu} A_{\nu]} +\\\nonumber &+& (v_2)^\mu_{AB} \, (\bar{\nu}_{L})_{A} \gamma^\nu (\nu_{L})_{B} \partial_{[\mu} A_{\nu]} +\\\nonumber &+& (v_3)^\mu_{AB} \, (\bar{\ell}_L)_{A} \gamma^\nu (\ell_{L})_{B} \partial_{[\mu} Z_{\nu]} +\\\nonumber &+&(v_4)^\mu_{AB} \, (\bar{\nu}_{L})_{A} \gamma^\nu (\nu_{L})_{B} \partial_{[\mu} Z_{\nu]} +
\\\nonumber &+& i g \, \rho^\mu_{AB} \, W^+_{[\mu} W^-_{\nu]} \times \\\nonumber
&&  \left( \, (\bar{\ell}_{L})_{A} \gamma^\nu (\ell_{L})_{B} - (\bar{\nu}_L)_{A} \gamma^\nu (\nu_{L})_{B} \, \right)\; + \\\nonumber
&+& \rho^\mu_{AB} \, (\bar{\ell}_{L})_{A} \gamma^\nu (\nu_{L})_{B} \nabla_{[\mu} W^-_{\nu]} + \text{h.c.}~~,\\
\end{eqnarray}

where we define for convenience
$v_{1\mu} = c_W \xi_{\mu} + s_W \rho_{\mu} \; $, $
v_{2\mu} =  c_W \xi_{\mu} - s_W \rho_{\mu} \; $, $
v_{3\mu} =  -s_W \xi_{\mu} + c_W \rho_{\mu} \; $, $
v_{4\mu} =  -s_W \xi_{\mu} - c_W \rho_{\mu} \;$. The Lorentz violation implemented into the Right lepton sector will be given by
\begin{equation}
\mathcal{L}_{LSV}^{Right} = \xi^\mu_{AB} \, (\bar{\ell}_{R})_{A} \gamma^\nu (\ell_{R})_{B} \, F_{\mu \nu} \; .
\end{equation}
Rewriting the above equation on the $\{ A_{\mu} \, , \, Z_{\mu} \}$ basis, we reach:
\begin{eqnarray}\nonumber
\mathcal{L}_{LSV}^{Right} &=& c_W \xi^\mu_{AB} \,(\bar{\ell}_{R})_{A} \gamma^\nu (\ell_{R})_{B} \, \partial_{[\mu}A_{\nu]}+\\
&&\hspace{-0.5cm}- s_W \xi^\mu_{AB} \, (\bar{\ell}_{R})_{A} \gamma^\nu (\ell_{R})_{B} \, \partial_{[\mu}Z_{\nu]} \; .
\end{eqnarray}

Let's now return to the full LSV Lagrangian. Using the definitions $(\ell_{L})_A = P_L\ell_A = \frac{1-\gamma_5}{2}\ell_A$ and $(\ell_{R})_A = P_R \ell_A = \frac{1+\gamma_5}{2} \ell_A$, $(\nu_L)_A = \nu_A$, with $\ell_A$ and $\nu_A$ Dirac spinors, we are able to rewrite the Lagrangian as follows:
\begin{eqnarray}\label{eqfinal} \nonumber
\mathcal{L}_{ LSV}&=&  \overline{\ell}_A \, \left( c_1^\mu \gamma^\nu + c_2^\mu \gamma^\nu \gamma_5 \, \right)_{AB} \, \ell_B \,  \partial_{[\mu}A_{\nu]} 
+\\\nonumber
&&\hspace{-0.5cm} + \frac{1}{2} \, (v_2)^\mu_{AB} \, \, \overline{\nu}_{A} \, \gamma^\nu \left(1-\gamma_5\right) \, \nu_{B} \,  \partial_{[\mu}A_{\nu]} +
 \\\nonumber
&&\hspace{-0.5cm}
+ \, \overline{\ell}_A \, \left( c_3^\mu \,  \gamma^\nu + c_4^\mu \gamma^\nu \gamma_5 \right)_{AB} \, \ell_B \, \partial_{[\mu}Z_{\nu]} 
+\\\nonumber
&& \hspace{-0.5cm}+ \frac{1}{2} \, (v_4)^\mu_{AB} \, \bar{\nu}_{A} \, \gamma^\nu \left(1-\gamma_5\right) \, \nu_{B} \, \partial_{[\mu}Z_{\nu]} + \\\nonumber
&& \hspace{-0.5cm}
+ \, \frac{1}{2} \, \rho^\mu_{AB} \, \overline{\ell}_A \, \gamma^\nu \, \left( 1-\gamma_5 \right) \, \nu_{B} \, \nabla_{[\mu} W^-_{\nu]}+
 \\ \nonumber
&&  \hspace{-0.5cm} + \, \frac{i}{2} \, g \, \rho^\mu_{AB} \, W^+_{[\mu} W^-_{\nu]} \times \\\nonumber &&\hspace{-0.5cm} \left[ \overline{\ell}_A \gamma^\nu \left(1-\gamma_5\right) \ell_B 
- \bar{\nu}_{A} \gamma^\nu \left( 1-\gamma_{5} \right)  \nu_{B} \right]+ \text{h.c.}~,\\
\end{eqnarray}
where $c_{1\mu} =  \frac{1}{2} (c_W \xi_{\mu} + \frac{1}{2} \, s_W \rho_{\mu})$, $c_{2\mu} = - \frac{1}{4} \, s_W \rho_{\mu}$, $c_{3\mu} = -\frac{1}{2}(  s_W \, \xi_{\mu} 
+ \frac{1}{2} \, c_W \rho_{\mu})$ e $c_{4\mu} =  \frac{1}{4} \, c_W \, \rho_{\mu}$ and the flavor indexes are omitted for simplicity. The above Lagrangian is a generalization of the model seen in Ref. \cite{elweakLSV1,nosso}, since in this work the flavor structure are take into account. 
\renewcommand{\arraystretch}{1.8}
\begin{table}[htb!]
\centering
\begin{tabular}{|c|c|}
\hline
\quad  {\bf Interaction}  \quad & \quad   {\bf Vertex}   \quad \\
\hline
\hline

 \quad $\gamma \, \ell_A \, \ell_B$  \quad &  \quad $q_\nu ( c_1^{[\mu} \gamma^{\nu]} + c_2^{[\mu} \gamma^{\nu]}\gamma_5 )_{AB}  $  \quad  \\
\hline

 \quad $\gamma \, \nu_{A} \, \nu_{B}$ \quad  &  \quad $\frac{1}{2} (v_2)_{AB}^{[\nu} \gamma^{\mu]} \frac{\left(1-\gamma_5\right)}{2} q_\nu$ \quad \\

\hline

 \quad $Z^0 \, \ell_A \, \ell_B$  \quad  &  \quad $q_\nu ( c_3^{[\mu} \gamma^{\nu]} + c_4^{[\mu} \gamma^{\nu]}\gamma_5 )_{AB} $  \quad  \\

\hline

\quad $Z^{0} \, \nu_{A} \, \nu_{B}$  \quad  &  \quad $\frac{1}{2} (v_4)_{AB}^{[\nu} \gamma^{\mu]} \frac{\left(1-\gamma_5\right)}{2} q_\nu$ \quad  \\

\hline

 \quad $W^{-} \, \ell_A \, \nu_{B}$ \quad   &  \quad $\frac{1}{\sqrt{2}} q_\nu (\rho)_{AB}^{[\nu} \gamma^{\mu]} \frac{\left(1-\gamma_5\right)}{2} $ \quad  \\

\hline

 \quad $W^{-} \, \gamma \, \ell_A \, \nu_{B}$  \quad & \quad $\frac{i e}{\sqrt{2}} (\rho)_{AB}^{[\nu} \gamma^{\mu]} \frac{\left(1-\gamma_5\right)}{2}$ \quad \\

\hline

 \quad $W^{-} \, Z^0 \, \ell_A \, \nu_{B}$ \quad &  \quad $\frac{i e}{\sqrt{2}}\cot \theta_W (\rho)_{AB}^{[\nu} \gamma^{\mu]} \frac{\left(1-\gamma_5\right)}{2}$ \quad \\

\hline

 \quad $W^{+} \, W^{-} \, \ell_A \, \ell_B$ \quad & \quad $\frac{i g}{2} (\rho)_{AB}^{[\nu} \gamma^{\mu]} \frac{\left(1-\gamma_5\right)}{2}$ \quad  \\

\hline

 \quad $W^{+} \, W^{-} \, \nu_{A} \, \nu_{B}$  \quad & \quad  $-\frac{i g}{2} (\rho)_{AB}^{[\nu} \gamma^{\mu]} \frac{\left(1-\gamma_5\right)}{2}$ \quad  \\

\hline
\end{tabular}
\caption{Vertex factors obtained from Eq.\eqref{eqfinal}. Here $q^\mu$ represents the  $A$, $W$ or $Z$ 4-momentum.} 
\label{vertexLV}
\end{table}

\paragraph*{}As we can see a novel coupling between the photon and the neutral current is generated. Going further, we can also see that a coupling between the electromagnetic current and the $ Z ^ 0 $ boson also arises.
A possible decay from Lorentz violation will be the so-called neutrino-free muon decay, $\mu \rightarrow e + \gamma$. This process is prohibited in the Standard Model, so that experimentally there are strong limits to this decay, with Branching Ratio $BR(\mu \rightarrow e + \gamma) < 4.2 \times 10^{-13} $ ($90\%$ C.L.) \cite{MEG}. In the same way as the lepton tau decays we have $BR(\tau \rightarrow e \gamma) < 3.3 \times 10^{-8}$ e $BR(\tau \rightarrow \mu \gamma) < 4.4 \times 10^{-8}$, both with $ 90 \% $ of confidence level \cite{MEG}.

\begin{figure}[htb!]
\centering
\includegraphics[height=5cm, angle = 0]{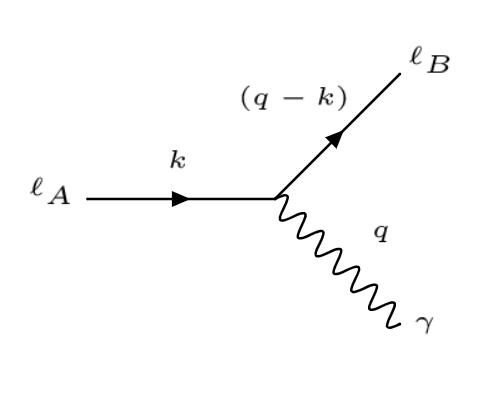}
\caption{ Hypothetical decay of lepton $ \ell_A$ into $ \ell _B$ and a photon due to LSV.} 
\label{FigLFV}
\end{figure}

\paragraph*{}From a momentum conservation perspective, the $ \ell_A \rightarrow \ell_B + \gamma $ decay can occur as long as the mass of the $ \ell_A $ lepton is greater than the $ \ell_B $ mass. However in the Standard Model this decay is prohibited, so we can use this decay to find bounds for the Lorentz violation parameters.
\paragraph*{} Directly, the scattering matrix which describes the decay of the Feynman diagram shown in Fig. \eqref{FigLFV} is given by:
\begin{eqnarray}\nonumber
\langle |M|^2 \rangle &=& Tr\left[ ( \slashed{q}- m_A )(\Gamma_{AB}^\mu) (\slashed{q}- \slashed{k} - m_B ) \bar{\Gamma}_{AB}^\nu  \right ]\\\nonumber && \times(-1)\sum_k\epsilon_{\mu,k} \epsilon^*_{\nu,k}~,
\end{eqnarray}

where $\Gamma_{AB}^\mu = q_\nu( c_1^{[\mu} \gamma^{\nu]} + c_2^{[\mu} \gamma^{\nu]}\gamma_5 )_{AB} $, $m_A > m_B$ and are no sum over $A,B$. Using the identity $\sum_k\epsilon_{\mu,k} \epsilon^*_{\nu,k} = \eta_{\mu \nu} - \frac{q_\mu q_\nu}{q^2}$ and calculating the traces over gamma matrices we get the following expression:
\begin{eqnarray}\nonumber
\langle |M|^2 \rangle &=& -12 m_A^2 \Big{[}-E_q (c_{20}^2 (m_A(1-y^2) + \\\nonumber &&\hspace{-1.5cm} +E_q ( y-3) y ) + 
   c_{10}^2 (m_A(1 -  y^2) + E_q y (3 + y))) + \\\nonumber
   && \hspace{-1.5cm} + c_{2}^0 \vec{c}_2 \cdot \vec{q} (m_A(1-y^2) + 2 E_q ( y-3) y ) + 
 \\ \nonumber && \hspace{-1.5cm}+ c_{1}^0 \vec{c}_1 \cdot \vec{q} (m_A(1- y^2) + 2 E_q y (3 + y))+\\
 && \hspace{-1.5cm} -y ((\vec{c}_2\cdot \vec{q})^2 ( y-3) + (\vec{c}_1 \cdot \vec{q})^2 (3 + y))  \Big{]}~~,
\end{eqnarray}
where $y = m_B/m_A$ and we hide the flavor indexes $c^\mu_{1,AB}$ and $c^\mu_{2,AB}$ for simplicity. With that, we can calculate the decay rate of this process and, using the rest frame of the lepton $ \ell_A $, is given by:
\begin{eqnarray}\nonumber
\Gamma(\ell_A \rightarrow \ell_B + \gamma) &=&\\\nonumber && \hspace{-2.5cm} =\frac{1}{2 (4 \pi)^2 m_A} \int d^3 \vec{q} \frac{\langle |\mathcal{M} |^2 \rangle}{E_q E_{k_j}} \delta( m_A - E_q - E_{k_B}) = \\\nonumber
 && \hspace{-2.5cm}=\frac{1}{2 (4 \pi)^2 m_A} \int d E_q E_q^2 \frac{\delta( m_A - E_q - E_{k_B})}{E_q E_{k_B}} \times \\
 && \times \int  d \Omega \langle |\mathcal{M} |^2 \rangle  ~~,
\end{eqnarray}
where
\begin{eqnarray}\nonumber
 \int d \Omega \langle |\mathcal{M} |^2 \rangle &\approx& -\frac{4}{3} E_q \pi \Big{(}   3 (c_{2}^0)^2  + 3 (c_{1}^0)^2  \Big{)}m_A + O(y)~~.
\end{eqnarray}
So:
\begin{eqnarray}\nonumber
\Gamma(\ell_A \rightarrow \ell_B + \gamma) &=&\\\nonumber
&& \hspace{-2.cm}=\Big{(}\frac{E_q}{8 \pi m_A E_{k_B}} \int \frac{d \Omega}{4 \pi} \langle |\mathcal{M} |^2 \rangle \Big{)}\Big{|}_{E_q = m_A-E_{k_B}}\\\nonumber &&\hspace{-2.cm} = \frac{3 \left((c_{1,AB}^0)^2 + (c_{2,AB}^0)^2 \right) m_A^3}{4 \pi} + O(y) .\\
\end{eqnarray}
 
\paragraph*{}Using the most recent measurements, we have that the mean life of the muon is given by $\tau_\mu = 3.3\times 10^{15} MeV^{-1}$, and for the tau-lepton $\tau_\tau = 4.4 \times 10^8 MeV^{-1}$ \cite{PDG}. Thus we have the Branching ratio for the muon decay is given by \cite{PDG}:
\begin{equation}
BR(\mu \rightarrow e + \gamma) = \frac{\Gamma(\mu \rightarrow e + \gamma)}{\tau_\mu^{-1}} < 4.2 \times 10^{-13}~~.
\end{equation}
 Then we reach the following bound: 
\begin{equation}
|\Delta_{12}| < 2.11 \times 10^{-17}  MeV^{-1}~~,
\end{equation}
where $\Delta_{12}^2 = \left((c_{1,12}^0)^2 + (c_{2,12}^0)^2 \right) $. Similarly  from the Branching Ratio $ BR(\tau \rightarrow \mu + \gamma) < 4.4 \times 10^{-8}$ \cite{PDG} we get the bound as follows:
\begin{equation}
|\Delta_{23}| < 2.7 \times 10^{-13}  MeV^{-1}~~,
\end{equation}
where $\Delta_{23}^2 = \left((c_{1,23}^0)^2 + (c_{2,23}^0)^2 \right) $. Finally, in the process $\tau \rightarrow e + \gamma$, we have the Branching ratio upper bound given by $ BR(\tau \rightarrow e + \gamma)  < 3.3 \times 10^{-8}$ \cite{PDG}. Thus we get the third limit which will be given by:
\begin{equation}
|\Delta_{13}| <  2.4 \times 10^{-13} MeV^{-1}~~,
\end{equation}
where $\Delta_{13}^2 = \left((c_{1,13}^0)^2 + (c_{2,13}^0)^2 \right)$. Rewriting the limits obtained in terms of the initial 4-vectors we have $ \Delta_{AB}^2 =  \frac{1}{8}(1 + cos 2 \theta_W)(\xi_{AB}^0)^2  + \frac{1}{4}  c_W s_W \rho_{AB}^0 \xi_{AB}^0 +
 \frac{1}{16}  (1 - cos 2 \theta_W)(\rho_{AB}^0)^2$, with no sum over flavor indexes. The bounds are grouped in table \ref{LFV1} and the region plots are shown in Fig. \ref{LFVbound}.  
\paragraph*{} The decays $\mu \rightarrow e + \gamma$ and $\tau \rightarrow \mu + \gamma$ are drastically suppressed in the SM; this is very important to distinguish between the neutrino flavors. Now, with LSV, these decays , though very tiny, can occur and they signal a very meaningful aspect of LSV in particle physics. 
 
\section{Final Comments}
\paragraph*{} In this work, we have focused on the analysis of the proposed new couplings (in flavor space) present in the model \cite{elweakLSV1} in the quark and lepton sectors. In the quark sector, based on the approximation given by eq. \eqref{sup1}, we assume that non-diagonal LSV components will affect the sector of the u-, c- and t-quarks, generating, for instance, $\Delta C \neq 0$ decays, i.e., $D^+ \rightarrow \pi^+  + \mu^+ + \mu^-$ . These events are highly suppressed in the SM, and this supports our approximation \eqref{sup1}. This type of decay is a good candidate to confirm if our assumption is solid. It would also allow us to find bounds on the spatial components of non-diagonal LSV parameters and could be a further step towards a more complete analysis of the model. We find that the CKM rotation could generate FCNC processes, even if the Lorentz violation parameters were diagonal in flavor space. We realize that, through the FCNC processes, only the spatial components or the 4-vector parameters, $\xi$ and $\rho$, contribute to the FCNC decays and we found limits between $|10^{-8} - 10^{-10}| \text{ MeV}^{-1}$ by considering the experimental FCNC bounds. 

\paragraph*{}As it can be seen in \cite{nosso}, we need to use the Sun-Centered Frame (SCF) in order to choose a reference system better than the Earth frame. So, in the SCF the modulus of the vector $\vec{V} = (V_x,V_y,V_z)$ is found:
\begin{eqnarray}\nonumber
|\vec{V}_{SCF}|^2 &=&  (\frac{1}{2}+\sin^2\chi)V_X^2 + (\frac{1}{2}+\cos^2\chi)V_Y^2 + \\
&& +V_Z^2   - 2 V_X V_Y \cos\chi \sin \chi  ~~,
\end{eqnarray} 

where $\chi$ is the co-latitude of the laboratory and $\vec{V}_{SCF} = (V_X,V_Y,V_Z)$. In the case of the LHC Collaborations, we have $\chi \approx 44º$.  

\begin{figure}[htb!]
\centering
\includegraphics[height=6cm, angle = 0]{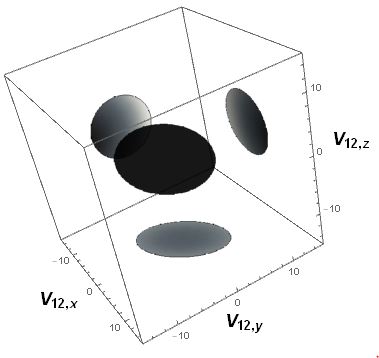}
\caption{ Plot of the allowed region in $\vec{V}_{12}$ parameter space, were $\vec{V}_{12} = \vec{\rho}_{11} - \vec{\rho}_{22} $. The figure is normalized in units of $10^{-10} MeV^{-1}$.}
\label{FCNC1bound}
\end{figure}

\begin{figure}[htb!]
\centering
\includegraphics[height=6cm, angle = 0]{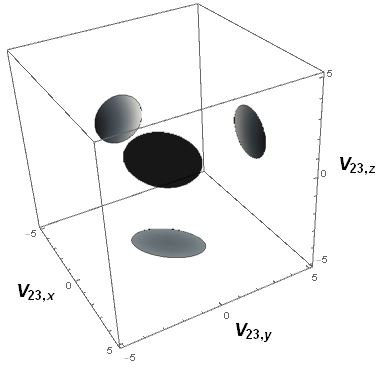}
\caption{ Plot of the allowed region in $\vec{V}_{23}$ parameter space, were $\vec{V}_{23} = \vec{\rho}_{22} - \vec{\rho}_{33} $.  The figure is normalized in units of $10^{-8} MeV^{-1}$.  }
\label{FCNC3bound}
\end{figure}

{\center 
\begin{table}[htb]
\centering
\begin{tabular}{|c|c|}
\hline

Decay & Bound ($MeV^{-1}$) \\ \hline \hline
\quad $K_L^0 \rightarrow \mu^+ + \mu^- $ \quad & \quad$|\Xi^{K^0}_{SCF} |  < 5.6 \times 10^{-10}$ \quad\\ \hline
\quad $ B^0 \rightarrow \mu^+ + \mu^- $ \quad & \quad  $|\Xi^{B^0}_{SCF}| < 4.2 \times 10^{-8}$ \quad  \\ \hline
\quad $B_s^0 \rightarrow \mu^+ + \mu^- $ \quad & \quad $ |\Xi^{B^0_s}_{SCF}| < 1.4 \times 10^{-8}$  \quad \\ \hline
\end{tabular}
\caption{Bounds for Lorentz-violating parameters from experimental limits of FCNC processes.}
\label{results1}
\end{table}
}
where \begin{eqnarray}\nonumber
\left(\Xi^{K^0}_{SCF}\right)^2 &=& 0.98 ~(\rho_{11,X}-\rho_{22,X})^2 +  \\\nonumber
&& \hspace{-1.4cm}+1.02~(\rho_{11,Y}-\rho_{22,Y})^2 + (\rho_{11,Z}-\rho_{22,Z})^2  + \\
&& \hspace{-1.4cm} - (\rho_{11,X}-\rho_{22,X}) (\rho_{11,Y}-\rho_{11,Y}) 
\end{eqnarray}
,
\begin{eqnarray}\nonumber
\left( \Xi^{B^0_s}_{SCF} \right)^2 &=& 0.98 ~(\rho_{22,X}-\rho_{33,X})^2 +  \\\nonumber
&& \hspace{-1.4cm} +1.02~(\rho_{22,Y}-\rho_{33,Y})^2 + (\rho_{22,Z}-\rho_{33,Z})^2   \\
&& \hspace{-1.4cm}- (\rho_{22,X}-\rho_{33,X}) (\rho_{22,Y}-\rho_{33,Y})
\end{eqnarray}  
\begin{eqnarray} \nonumber
\left( \Xi^{B^0}_{SCF} \right)^2 &=&  0.98 \left( W_{X}^2- 0.12 ~(\rho_{11,X}-\rho_{33,X})^2 \right) +  \\\nonumber && \hspace{-1.4cm}+ 1.02 \left( W_{Y}^2- 0.12~(\rho_{11,Y}-\rho_{33,Y})^2 \right) + \\\nonumber && \hspace{-1.4cm}+\left( W_{Z}^2- 0.12 ( \rho_{11,Z}-\rho_{33,Z})^2 \right)  + \\\nonumber &&\hspace{-1.4cm} -\left( W_{X} W_{Y} - 0.12~(\rho_{11,X}-\rho_{33,X}) (\rho_{11,Y}-\rho_{33,Y}) \right) ~,\\
\end{eqnarray} 
where $\vec{W} = (1- \rho)\vec{\rho}_{11,SCF} - \vec{\rho}_{22,SCF} + \rho \vec{\rho}_{33,SCF} =  (0.98\vec{\rho}_{11} - \vec{\rho}_{22} + 0.12 \vec{\rho}_{33})_{SCF}$, $\rho \approx 0.12$ and $\eta \approx 0.35$ \cite{SM4}. As seen above, the limits are not simple, specially the limit from the $B^0$-decay. Besides the complexity of the bound obtained from the $B^0$-decay, the bound reached from the $K^0$- and the $B^0_s$-decays can be visualized as Ellipsoids (see Fig. \eqref{FCNC1bound} and \eqref{FCNC3bound}). Thus, we can state that the modulus of $|\vec{\rho}_{11} - \vec{\rho}_{22}|_{SCF}$ and $|\vec{\rho}_{22} - \vec{\rho}_{33}|_{SCF}$ are approximately below $10^{-9} MeV^{-1}$ and $10^{-7}MeV^{-1}$, respectively.  

\paragraph*{} As we have shown, the FCNC decays impose bounds on the spatial components of the non-diagonal(in flavor space) LSV parameters $\rho^\mu_{AB}$. On the other hand, the lepton sector gives us bounds that depend on the time components $\rho^0_{AB}$.  Analyzing from the SCF perspective, we have $\rho^0_{AB} \approx \rho^T_{AB}$, where $\rho^T_{AB}$ is the time component of $\rho^\mu_{AB,SCF}$. Therefore, we find that bounds between ($|10^{-12} - 10^{-16}| \text{ MeV}^{-1}$) hold for the non-diagonal components of the LSV parameters through the bounds in the expression below:
{\small
\begin{equation}
|\Delta_{AB}| = \sqrt {0.19\, (\xi^T_{AB})^2 + 0.10\, \xi^T_{AB}\, \rho^T_{AB} + 0.03\, (\rho^T_{AB})^2} ,
\end{equation}
}
where we have used the lepton flavor violation branching ratios. The results are shown in table \eqref{LFV1}. It is worthy to highlight that our bounds on the lepton sector are, when compared with our weakest bound, five times  more accurate than the bound reached in Ref. \cite{nosso}  (where $\rho^T < 8 \times 10^{-7} MeV^{-1}$ is found, also from weak decays). A final comment should be done in order to shed some light on the QED sub-sector, where the present LSV coupling comes from. In table \eqref{vertexLV}, we see that the coupling parameter given by $c_1^\mu  = \frac{1}{2}\cos \theta_W \xi^\mu + \frac{1}{4} \sin \theta_W \rho^\mu$ acts on the electron-photon sector, and this coupling can be compared with the coupling used in \cite{nosso2}. In the aforementioned paper, some bounds are fixed through QED processes and the best bounds found are given by $c_1^T <  10^{-6} MeV^{-1}$; the allowed region given by this limit is depicted in Fig. \ref{LFVbound}. By analyzing it, we can state that our choice of considering the LFV sector gives us stronger bounds, 7 order of magnitude, in the case of B-mesons, up to 10 orders of magnitude in the Kaon decay case.  
\paragraph*{}Finally, we can see that, in the electron sector, the interaction Lagrangian given by $\bar{\ell}_1 (c_1^\mu \gamma^\nu + c_2^\mu \gamma^\nu \gamma_5)_{11} \ell_1 \partial_{[\mu} A_{\nu]} \supset c_2^0 \langle \vec{\sigma} \rangle \cdot \vec{E} $, generates an Electric Dipole Moment (EDM) to the electron (in fact for all fermions). Since the ACME experiment reveals that the magnitude of the electron's EDM has the upper bound $d_e < 8.7 \times 10^{-29} e \cdot cm  = 1.3 \times 10^{- 16} MeV^{-1} $ \cite{EDM}, we attain the following bound: 

\begin{equation}
\rho^T_{11} < 2.5 \times 10^{- 16} MeV^{-1} 
\end{equation}

\begin{widetext}
{\center
\begin{figure}[H]
\begin{subfigure}{1.\textwidth}
  \centering
\includegraphics[height=10cm, angle = 0]{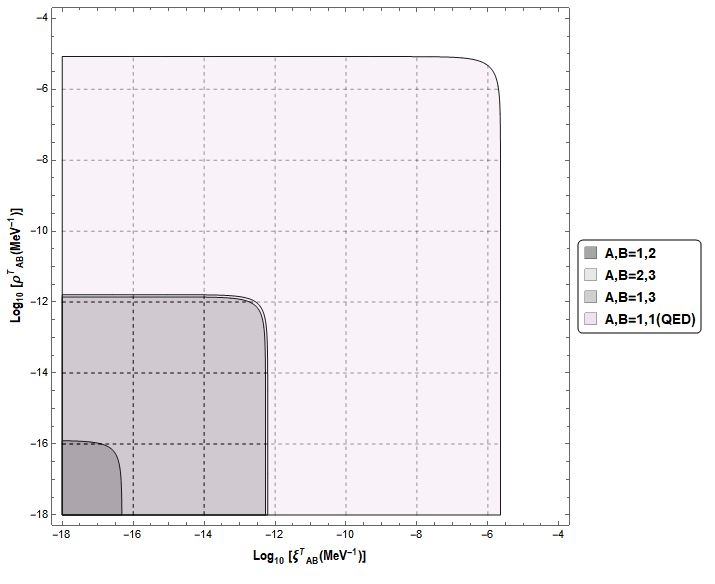}
\caption{ Plot of the allowed regions in $\xi^T_{AB} \times \rho^T_{AB}$ parameter space. Here $A,B = 1,2$ refer to $\mu \rightarrow e + \gamma$ decay, $A,B = 2,3$ refers to $\tau \rightarrow \mu + \gamma$ decay, $A,B = 1,3$ refers to $\tau \rightarrow e + \gamma$ decay and the $A,B= 1,1$ case refers to recent QED-based bounds \cite{nosso2}. We use $\theta_W = \arccos (80/91)$. }
\label{LFVbound}
\end{subfigure}
\newline
\begin{subfigure}{1.\textwidth}
\centering
\includegraphics[height=10cm, angle = 0]{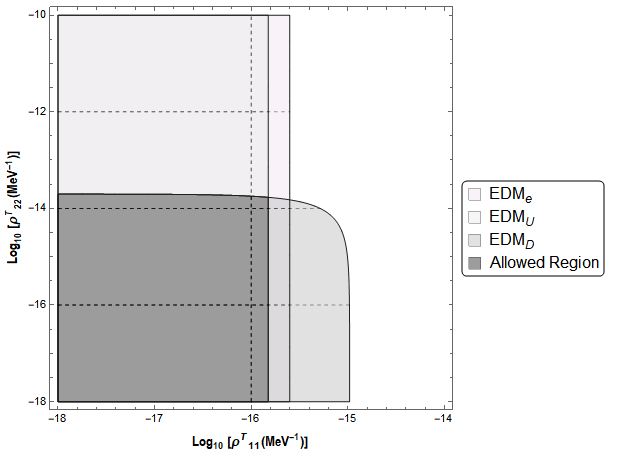}
\caption{  Plot of the allowed region in $\rho_{11}^T \times \rho_{22}^T$ parameter space from EDM bounds.   }
\label{EDMbound}
\end{subfigure}
\end{figure}
}
\end{widetext}
\paragraph*{} Besides the strong limit above, no other lepton-EDM can be used to impose limits on LSV parameters, since neither muon nor tau EDM experiments have been carried out so far. In the quark sector, we proceed in a similar way. By taking the best bounds on the up- and down-quarks EDMs, obtained through the proton and neutron EDMs \cite{qEDM}, we find the bounds $d_u< 1.1 x 10^{-27} e . cm = 1.5 \times 10^{-15} MeV^{-1}$ and $d_d< 6.5 \times 10^{-28} e.cm = 9.9 \times 10^{-16} MeV^{-1}$, respectively. Thus, the following bounds can be set:
\begin{equation}
\rho^T_{11} < 1.5 \times 10^{-16} MeV^{-1};
\end{equation}
and
\begin{equation}
\rho^T_{11} + 0.05(\rho^T_{22} - \rho^T_{11}) < 9.9 \times 10^{-16} MeV^{-1}  
\end{equation}
\paragraph*{}As it can be seen, the limit obtained by the u-quark EDM gives us a bound on $\rho_{11}^T$ a little better than in the case we adopt the electron EDM. By taking the d-quark EDM, we can find a bound on the $\rho_{22}^T$ LSV parameter. These limits may be seen in table \eqref{LFV1} and in fig. \ref{EDMbound}. Due to experimental limitations in measuring the EDMs of other SM particles, there are no other possible EDM bounds until the moment. 
\paragraph*{}As a step forward, a more complete analysis of the neutrino sector should be studied, particularly
 the neutrino-photon sector. The new coupling shown in table \eqref{LFVbound} could be limited by astrophysical measurements and could bring interesting insights about the relation between LSV and flavor physics in neutrino interactions.



\begin{table}[H]
\centering
\begin{tabular}{|c|c|}
\hline
\quad Event \quad &  \quad Bound ($MeV^{-1}$) \\ \hline \hline
\quad $\mu \rightarrow e + \gamma$ \quad \quad & \quad $|\Delta_{12} | < 2.1\times 10^{-17}$ \quad \\ \hline
\quad $\tau \rightarrow \mu + \gamma$ \quad \quad & \quad $ |\Delta_{23}| < 2.7 \times 10^{-13}$ \quad \\ \hline
\quad $\tau \rightarrow e + \gamma$ \quad \quad &  \quad $ |\Delta_{13} | < 2.4 \times 10^{-13}$\quad  \\ \hline
$EDM_e$ \quad \quad & \quad $\rho_{11}^T < 2.5 \times 10^{-16} $\\ \hline
$EDM_U$ \quad \quad & \quad $\rho_{11}^T < 1.5 \times 10^{-16} $\\ \hline
$EDM_D$ \quad \quad & \quad $\rho^T_{11} + 0.05(\rho^T_{22} - \rho^T_{11}) < 9.9 \times 10^{-16} $\\ \hline
\end{tabular}
\caption{ Limits for Lorentz violation parameters from the experimental limits of the lepton LFV sector.  }
\label{LFV1}
\end{table}




\begin{acknowledgements}

This work was funded by the Brazilian National Council for Scientific and Technological Development (CNPq).  We are thankful for the reviewer comments and suggestions. 

\end{acknowledgements}

\bibliographystyle{is-unsrt}
\bibliography{biblio}

\begin{thebibliography}{10}

\bibitem{intro1}
V~Alan Kosteleck{\`y} and Stuart Samuel.
\newblock Spontaneous breaking of lorentz symmetry in string theory.
\newblock {\em Physical Review D}, 39\penalty0 (2):\penalty0 683, 1989.

\bibitem{intro2}
Irina Mocioiu, Maxim Pospelov, and Radu Roiban.
\newblock Breaking cpt by mixed noncommutativity.
\newblock {\em Physical Review D}, 65\penalty0 (10):\penalty0 107702, 2002.

\bibitem{intro3}
Stefano Liberati.
\newblock Tests of lorentz invariance: a 2013 update.
\newblock {\em Classical and Quantum Gravity}, 30\penalty0 (13):\penalty0
  133001, 2013.

\bibitem{intro4}
David Mattingly.
\newblock Modern tests of lorentz invariance.
\newblock {\em Living Reviews in relativity}, 8\penalty0 (1):\penalty0 5, 2005.

\bibitem{intro5}
C~Adam and Frans~R Klinkhamer.
\newblock Causality and cpt violation from an abelian chern--simons-like term.
\newblock {\em Nuclear Physics B}, 607\penalty0 (1-2):\penalty0 247--267, 2001.

\bibitem{intro6}
Rodolfo Casana, Manoel~M Ferreira~Jr, and Carlos~EH Santos.
\newblock Classical solutions for the carroll-field-jackiw-proca
  electrodynamics.
\newblock {\em Physical Review D}, 78\penalty0 (2):\penalty0 025030, 2008.

\bibitem{intro7}
AP~Baeta Scarpelli, Humberto Belich, JL~Boldo, and JA~Helayel-Neto.
\newblock Aspects of causality and unitarity and comments on vortexlike
  configurations in an abelian model with a lorentz-breaking term.
\newblock {\em Physical Review D}, 67\penalty0 (8):\penalty0 085021, 2003.

\bibitem{intro8}
Don Colladay and V~Alan Kosteleck{\`y}.
\newblock Lorentz-violating extension of the standard model.
\newblock {\em Physical Review D}, 58\penalty0 (11):\penalty0 116002, 1998.

\bibitem{intro9}
V~Alan Kosteleck{\`y} and Matthew Mewes.
\newblock Signals for lorentz violation in electrodynamics.
\newblock {\em Physical Review D}, 66\penalty0 (5):\penalty0 056005, 2002.

\bibitem{intro10}
Yunhua Ding and V~Alan Kosteleck{\`y}.
\newblock Lorentz-violating spinor electrodynamics and penning traps.
\newblock {\em Physical Review D}, 94\penalty0 (5):\penalty0 056008, 2016.

\bibitem{intro11}
V~Alan Kosteleck{\`y} and Zonghao Li.
\newblock Gauge field theories with lorentz-violating operators of arbitrary
  dimension.
\newblock {\em Physical Review D}, 99\penalty0 (5):\penalty0 056016, 2019.

\bibitem{elweakLSV1}
Victor~E Mouchrek-Santos and Manoel~M Ferreira~Jr.
\newblock Repercussions of dimension five nonminimal couplings in the
  electroweak model.
\newblock In {\em Journal of Physics: Conference Series}, volume 952, page
  012019. IOP Publishing, 2018.

\bibitem{SM1}
Aneesh~V Manohar and Mark~B Wise.
\newblock {\em Heavy quark physics}, volume~10.
\newblock Cambridge university press, 2000.

\bibitem{MEG}
E~Ripiccini, MEG Collaboration, et~al.
\newblock New result from the meg experiment at psi and the meg upgrade.
\newblock {\em Nuclear and Particle Physics Proceedings}, 260:\penalty0
  147--150, 2015.

\bibitem{kostmesons}
Benjamin~R Edwards and V~Alan Kosteleck{\`y}.
\newblock Searching for cpt violation with neutral-meson oscillations.
\newblock {\em Physics Letters B}, 795:\penalty0 620--626, 2019.

\bibitem{bsmesonLHC}
Roel Aaij, C~Abell{\'a}n Beteta, B~Adeva, M~Adinolfi, Z~Ajaltouni, S~Akar,
  J~Albrecht, F~Alessio, M~Alexander, S~Ali, et~al.
\newblock Search for violations of lorentz invariance and c p t symmetry in b
  (s) 0 mixing.
\newblock {\em Physical review letters}, 116\penalty0 (24):\penalty0 241601,
  2016.

\bibitem{SM4}
Gerhard Buchalla and Andrzej~J Buras.
\newblock Qcd corrections to rare k-and b-decays for arbitrary top quark mass.
\newblock {\em Nuclear Physics B}, 400\penalty0 (1-3):\penalty0 225--239, 1993.

\bibitem{PDG}
Masaharu Tanabashi, K~Hagiwara, K~Hikasa, K~Nakamura, Y~Sumino, F~Takahashi,
  J~Tanaka, K~Agashe, G~Aielli, C~Amsler, et~al.
\newblock Review of particle physics.
\newblock {\em Physical Review D}, 98\penalty0 (3):\penalty0 030001, 2018.

\bibitem{nosso}
YMP Gomes, PC~Malta, and MJ~Neves.
\newblock Testing lorentz-symmetry violation via electroweak decays.
\newblock {\em arXiv preprint arXiv:1909.10398}, 2019.

\bibitem{nosso2}
et~al. de~Brito, G.~P.
\newblock Lorentz violation in simple qed processes.
\newblock {\em Physical Review D 94.5 : 056005}, 2016.

\bibitem{EDM}
et~al. Baron, Jacob.
\newblock Order of magnitude smaller limit on the electric dipole moment of the
  electron.
\newblock {\em Science 343.6168 : 269-272}, 2014.

\bibitem{qEDM}
Zhiwen~Zhao Liu, Tianbo and Haiyan Gao.
\newblock Experimental constraint on quark electric dipole moments.
\newblock {\em Physical Review D 97.7 : 074018}, 2018.

\end{thebibliography}

\end{document}